\documentstyle[12pt] {article}
\begin {document}
\parindent=15pt
\begin{center}
\vskip 1.5 truecm
{\bf IMPACT PARAMETER DEPENDENCES OF THE NUMBER OF INTERACTING NUCLEONS
AND MEAN MULTIPLICITIES IN HIGH ENERGY HEAVY ION INTERACTIONS}\\
\vspace{.5cm}
C.Pajares \\
\vspace{.5cm}
Departamento de Fisica de Particulas, Universidade de Santiago de
Compostela, 15706-Santiago de Compostela, Spain \\
\vspace{.5cm}
and \\
\vspace{.5cm}
Yu.M.Shabelski \\
\vspace{.5cm}
Petersburg Nuclear Physics Institute, \\
Gatchina, St.Petersburg 188350 Russia \\
\end{center}
\vspace{1cm}
\begin{abstract}
We consider the dependences of the average number of interacting
nucleons in high energy heavy ion collisions on the impact parameter
in two cases, when the colliding nuclei have equal atomic weights, and
when one nucleus is significantly more heavy in comparison with the
second one. We argue that in the case of trigger of some rare event
(say, $J/\psi$, or $\Upsilon$ production) the multiplicity of the
secondaries can change several times for minimum bias sample, but it
should be stable in the case of central events.
\end{abstract}
\vspace{3cm}

\newpage

\section{Introduction}
The correlation of mean multiplicity with some trigger is the
important problem of high energy heavy ion physics. For example,
the $J/\psi$ suppression can be explained at least partially
\cite{CKKG,ACF} by their
interactions with co-moving hadrons, and for the numerical
calculations we should know the multiplicity of comovers namely
in events with $J/\psi$ production.

In the present paper we give some results for the dependences of
the number of interacting nucleons and the multiplicity of produced
secondaries on the impact parameter. These results are based practically
only on geometry, and do not depend on the model of interaction. In the
case of minimum bias interactions the dispersion of the distribution on
the number of interacting nucleons (which is similar to the
distributions on the transverse energy, multiplicity of secondaries,
etc.) is very large. This allows in principle to have a significant
dependence of some characteristic of the interaction, say, mean
multiplicity of the secondaries, on the used trigger. On the other hand,
in the case of central collisions the discussed dispersion is small that
should result in weak dependence on any trigger.

We consider the high energy nucleus-nucleus collision as a superposition
of the independent nucleon-nucleon interactions. So our results can be
considered also as a test for search the quark-gluon plasma formation.
In the case of any collective interactions, including the case of
quark-gluon plasma formation we can not see any reason for existance the
discussed ratios. We present an estimation of possible violation which
is based on the quark-gluon string fusion calculations.

\section{Distributions on the number of interacting nucleons
for different impact parameters}

Let us consider the events with secondary hadron production in
nuclei A and B minimum bias collisions. In this case the average number
of inelastically interacting nucleons of a nucleus A is equal \cite{BBC}
to

\begin{equation}
<N_A>_{m.b.} = \frac{A \sigma^{prod}_{NB}}{\sigma^{prod}_{AB}} \;.
\end{equation}

If both nuclei, A and B are heavy enough, the production cross sections
of nucleon-nucleus and nucleus-nucleus collisions can be written as

\begin{equation}
\sigma^{prod}_{NB} = \pi R_B^2 \;,
\end{equation}
and
\begin{equation}
\sigma^{prod}_{AB} = \pi (R_A + R_B)^2 \;.
\end{equation}
It is evident that in the case of equal nuclei, A = B,
\begin{equation}
<N_A>_{m.b.} = A/4 \;.
\end{equation}
So in the case of minimum bias events the average number of
interacting nucleons should be four times smaller than in the
case of central collisions, where $<N_A>_c \approx A$.

For the calculation of the distribution over the number of
inelastically ineracting nucleons of A nucleus we will use the
rigid target approximation \cite{Alk,VT,PR}, which gives the
probability of $N_A$ nucleons interaction as \cite{GCSh,BSh}

\begin{equation}
V(N_A) = \frac{1}{\sigma^{prod}_{AB}} \frac{A!}{(A-N_A)! N_A!}
\int d^2 b [I(b)]^{A-N_A} [1-I(b)]^{N_A} \;,
\end{equation}
where
\begin{equation}
I(b) = \frac{1}{A} \int d^2 b_1 T_A(b_1-b)
\exp {- \sigma^{inel}_{NN}T_B(b_1)]} \;,
\end{equation}
\begin{equation}
T_A(b) = A \int dz \rho (b,z) \;.
\end{equation}

Eq. (5) is written for minimum bias events. In the case of events
for some interval of impact parameter $b$ values, the integration
in Eq. (5) should be fulfilled by this interval,
$b_{min} < b < b_{max}$. In particular, in the case of central
collisions the integration should be performed with the condition
$b \leq b_0$, and $b_0 \ll R_A$.

The calculated results for averaged values of the number of inelastic
interacting nucleons of the projectile nucleus, $<N_{in}>$ are presented
in Fig. 1, as the functions of impact parameter $b$ for the cases of
$Pb-Pb$ collisions at three different energies (we define $\sqrt{s}$ =
$\sqrt{s_{NN}}$ as the c.m.energy for one nucleon-nucleon pair), and for
$S-U$ collisions at $\sqrt{s_{NN}}$ = 20 GeV. One can see very
weak energy dependence of these distributions on the initial energy,
which appears in our approach only due to the energy dependence of
$\sigma_{NN}^{inel}$.

In the case of the collisions of equal heavy ions ($Pb-Pb$ in our case)
at zero impact parameter, about 6\% of nucleons from every nucleus do
not interact inelastically at energy $\sqrt{s_{NN}}$ = 18 GeV. More
accurate, we obtain on the average 11.8 non-interacting nucleons at this
energy, that is in agreement with the value $13 \pm 2$ nucleons
\cite{Alb}, based on the VENUS 4.12 \cite{Kla} model prediction. The
number of non-interacting nucleons decreases to the value about 3\% at
$\sqrt{s_{NN}}$ = 5.5 TeV. This is connected with the fact that the
nucleons at the periphery of one nucleus, which are overlapped with the
region of small density of nuclear matter at the periphery of another
nucleus, have large probability to penetrate without inelastic
interaction. It is clear that this probability decrease with increase of
$\sigma_{NN}^{inel}$, that results in the presented energy dependence.
The value of $<\!N_{in}\!>$ decreases with increase of the impact
parameter because even at small $b \neq 0$ some regions of colliding
ions are not overlapping.

In the case of different ion collisions, say $S-U$, at small impact
parameters all nucleons of light nucleus go throw the regions of
relatively high nuclear matter density of heavy nucleus, so
practically all these nucleons interact inelastically. For the case of
$S-U$ interactions at $\sqrt{s_{NN}}$ = 20 GeV it is valid for
$b < 2\div 3$ fm.

It is interesting to consider the distributions on the number of
inelastically interacting nucleons at different impact parameters.
The calculated probabili\-ties to find the given numbers of
inelastically interacting nucleons for the case of minimum bias $Pb-Pb$
events are presented in Fig. 2a. The average value, $<N_{in}>$ = 50.4
is in reasonable agreement with Eq. (4). The disagreement of the order
of 3\% can be connected with different values of effective nuclear radii
in Eqs. (2) and (3). The dispersion of the distribution on $N_{in}$ is
very large.

The results of the same calculations for different regions of impact
parame\-ters are presented in Fig. 2b, where we compare the cases of the
central ($b < 1$ fm), peripheral (12 fm $< b <$ 13 fm) and intermediate
(6 fm $< b <$ 7 fm) collisions. One can see that the dispersions of all
these distributions are many times smaller in comparison with the
minimum bias case, Fig. 2a.

In the cases of the central and peripheral interactions, the
distributions over $N_{in}$ are significantly more narrow than in the
intermediate case. The reason is that in the case of central collision
the number of nucleons at the periphery on one nucleus, which have the
probabilities to interact or not of the same order, is small enough.
In the case of very peripheral collision the total number of nucleons
which can interact is small. However, in the intermediate case the
comparatively large number of nucleons of one nucleus go via
peripheral region of another nucleus with small nuclear matter density,
and every of these nucleons can interacts or not.

\section{Ratio of secondary hadron multiplicities in the central
and minimum bias heavy ion collisions}

Let us consider now the multiplicity of the produced secondaries in the
central region. First of all, it should be proportional to the number of
interacting nucleons of projectile nucleus. It should depends also on
the average number, $<\! \nu_{NB} \!>$, of inelastic interactions of
every projectile nucleon with the target nucleus. At asymptotically
high energies the mean multiplicity of secondaries produced in
nucleon-nucleus collision should be proportional to $<\! \nu\! >$
\cite{Sh1,CSTT}. As was shown in \cite{Sh}, the average number of
interactions in the case of central nucleon-nucleus collisions,
$<\! \nu \!>_c$, is approximately 1.5 times larger than in the case of
minimum bias nucleon-nucleus collisions, $<\! \nu \!>_{m.b.}$. It means
that the mean multiplicity of any secondaries in the central heavy ion
collisions (with A = B), $<\! n\!>_c$ should be approximately 6 times
larger than in the case of minimum bias collisions of the same nuclei,
$<\! n \!>_{m.b.}$, $<\! n \!>_c \approx 6 <\! n \!>_{m.b.}$. Of course,
this estimations is valid only for secondaries in the central region of
inclusive spectra.

There exist several corrections to the obtained result. At existing
fixed target energies the multiplicity of secondaries is proportional
not to $<\! \nu \!>$, but to $\frac{1 + <\nu>}{2}$ \cite{CSTT,CCHT}.
For heavy nuclei the values of $<\! \nu \!>_{m.b.}$ are about
$3 \div 4$. It means, that the $<\nu_{NB}>_c$ to $<\nu_{NB}>_{m.b.}$
ratio equal to 1.5 will results in enhancement factor about 1.4 for the
multiplicity of secondaries. More important correction comes from the
fact that in the case of central collision of two nuclei with the same
atomic weights, only a part of projectile nucleons can interact with the
central region of the target nucleus. It decrease the discussed
enhancement factor to, say, 1.2. As it was presented in the previous
Sect., even in central collisions (with zero impact parameter) of equal
heavy nuclei, several percents of projectile nucleons do not interact
with the target because they are moving through the diffusive region of
the target nucleus with very small density of nuclear matter.

As a result we can estimate our prediction
\begin{equation}
<\! n \!>_c \sim 4.5 <\! n \!>_{m.b.} \;.
\end {equation}

In the case of quark-gluon plasma formation or some another collective
effects we can not see the reason for such predictions. For example,
the calculation of $<n>_c$ and $<n>_{m.b.}$ with account the string
fusion effect \cite{USC} violate Eq. (8) on the level of 40\% for the
case of $Au-Au$ collisions at RHIC energies.

Moreover, in the conventional approach considered here, we obtain the
prediction of Eq. (8) for any sort of secondaries including pions,
kaons, $J/\psi$, Drell-Yan pairs, direct photons, etc. Let us imagine
that the quark-gluon plasma formation is possible only at comparatively
small impact parameters (i.e. in the central interactions). In this case
Eq. (8) can be strongly violated, say, for direct photons and, possibly,
for light mass Drell-Yan pairs, due to the additional contribution to
their multiplicity in the central events via thermal mechanism. At the
same time, Eq. (8) can be valid, say, for pions, if the most part of
them is produced at the late stage of the process, after decay of the
plasma state. So the violation of Eq. (8) for the particles which can
be emitted from the plasma state should be considered as a signal for
quark-gluon plasma formation. Of course, the effects of final state
interactions, etc. should be accounted for in such test.

It was shown in Ref. \cite{DPS} that the main contribution to the
dispersion of multiplicity distribution in the case of heavy ion
collisions comes from the dispersion in the number of nucleon-nucleon
interactions. The last number is in strong correlation
with the value of impact parameter.

For the normalized dispersion $D/<\! n \!>$, where
$D^2 = <\! n^2 \!> - <\! n \!>^2$
we have \cite{DPS}
\begin{equation}
\frac{D^2}{<\! n \!>^2} =
\frac{\nu_{AB}^2 - <\nu_{AB}>^2}{<\nu_{AB}>^2}
+ \frac{1}{<\nu_{AB}>} \frac{d^2}{\overline{n}^2} \;,
\end {equation}
where $<\nu_{AB}> = <\! N_A \!> \cdot <\nu_{NB}>$ is the average number
of nucleon-nucleon interactions in nucleus-nucleus collision,
$\overline{n}$ and $d$ are the average multiplicity and the dispersion
in one nucleon-nucleon collision.

In the case of heavy ion collisions $<\nu_{AB}> \sim 10^2 - 10^3$,
so the second term in the right hand side of Eq. (9) becomes negligible
\cite{DPS}, and the first term, which is the relative dispersion in the
number of nucleon-nucleon interactions, dominates. In the case of
minimum bias A-B interaction the last dispersion is comparatively large
due to large dispersion in the distribution on $N_A$, see Fig. 2a. So in
the case of some trigger (say, $J/\psi$ production) without fixing of
the impact parameter, the multiplicity of secondaries can change
significantly in comparison with its average value. In the case of some
narrow region of impact parameters the dispersion in the distribution on
$N_A$ is many times smaller, as one can see in Fig. 2b, especially in
the case of central collisions. The dispersion in the number of
inelastic interactions of one projectile nucleon with the target
nucleus, $\nu_{NB}$, should be the same or slightly smaller in
comparison with the minimum bias case. So the dispersion in the
multiplicity of secondaries can not be large. It means that any trigger
can not change significantly the average multiplicity of secondaries in
the central heavy ion collisions, even if this trigger strongly
influents on the multiplicity in the nucleon-nucleon interaction.

\section{Conclusions}

We calculated the distributions on the number of interacting nucleons
in heavy ion collisions as the functions of impact parameters. The
dispersions of these distributions are very small for the central and
very peripheral interactions and significantly larger for
intermediate values of impact parame\-ters.

We estimated also the ratio of mean multiplicities of secondaries in
minimum bias and central collisions, which can be used for search of
quark-gluon plasma formation. We presented that in the case of 
central collisions any trigger can not change significantly (say,
more than 10-15\%) the average multiplicity. This fact can be used
experimentally to distinguish collective effects on $J/\psi$ production
like quark-gluon plasma from more conventional machanisms.

In conclusion we express our gratitude to A.Capella and A.Kaidalov
for useful discussions. We thank the Direcci\'on General de
Pol\'{\i}tica Cient\'{\i}fica and the CICYT of Spain for financial
support. The paper was also supported in part by grant NATO OUTR.LG
971390.

\newpage

\begin{center}
{\bf Figure captions}\\
\end{center}

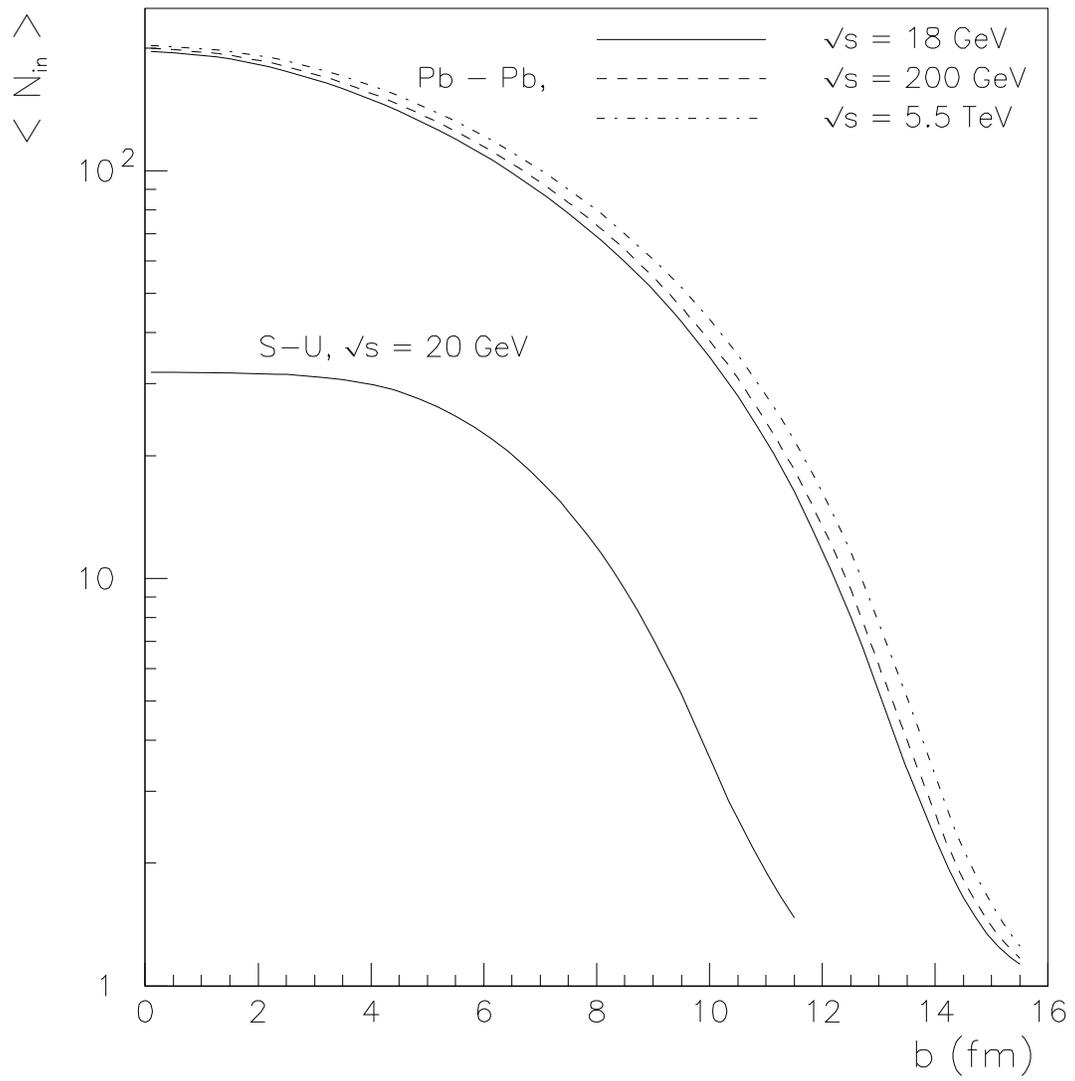
Fig. 1. Average numbers of inelastically interected nucleons in
$Pb-Pb$ and $S-U$ collisions at different energies as the functions of
impact parameter.

Fig. 2. Distributions on the numbers of inelastically interected
nucleons in $Pb-Pb$ collisions at $\sqrt{s_{NN}} = 18$ GeV for minimum
bias (a) interactions and for different regions of impact parameter (b).

\end {document}
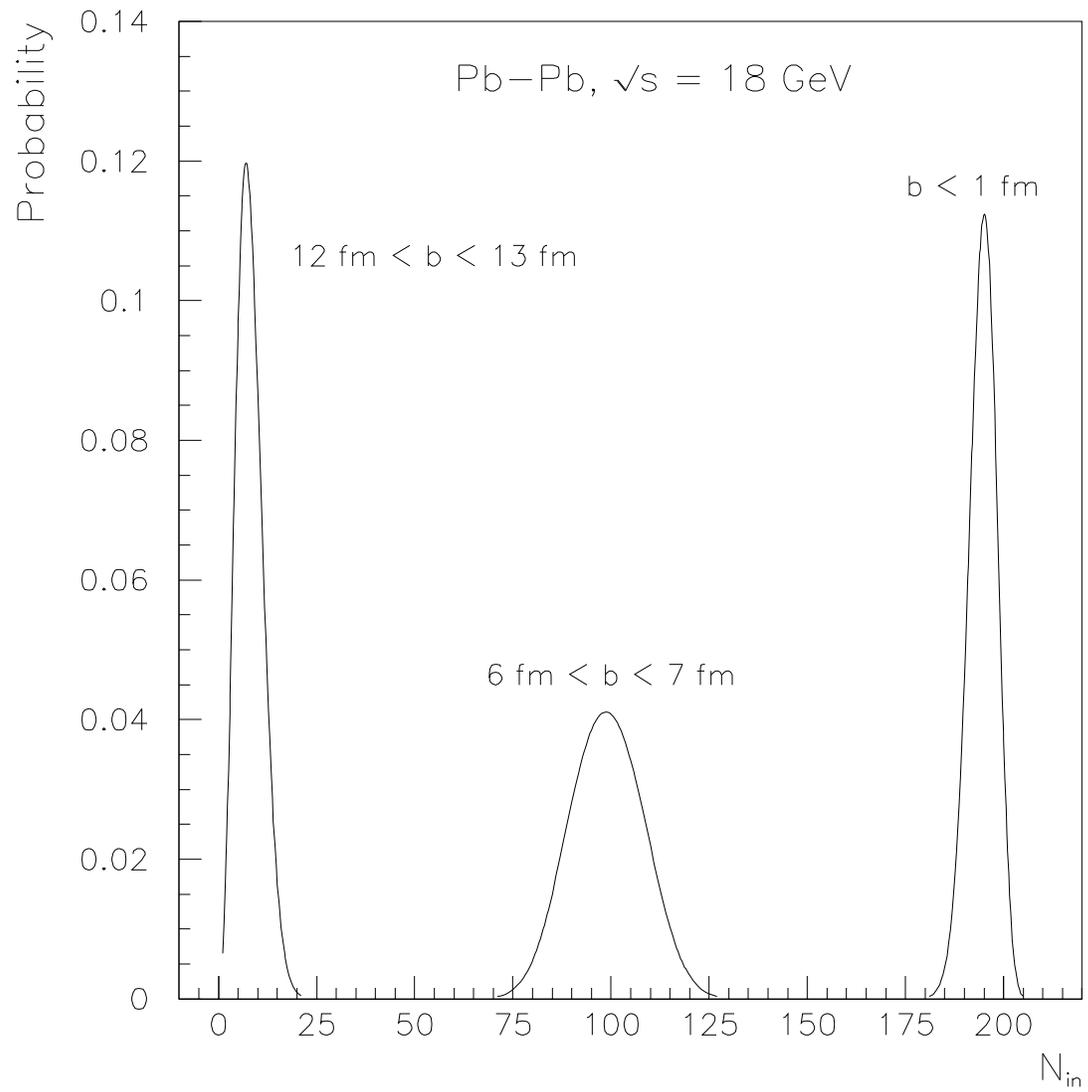
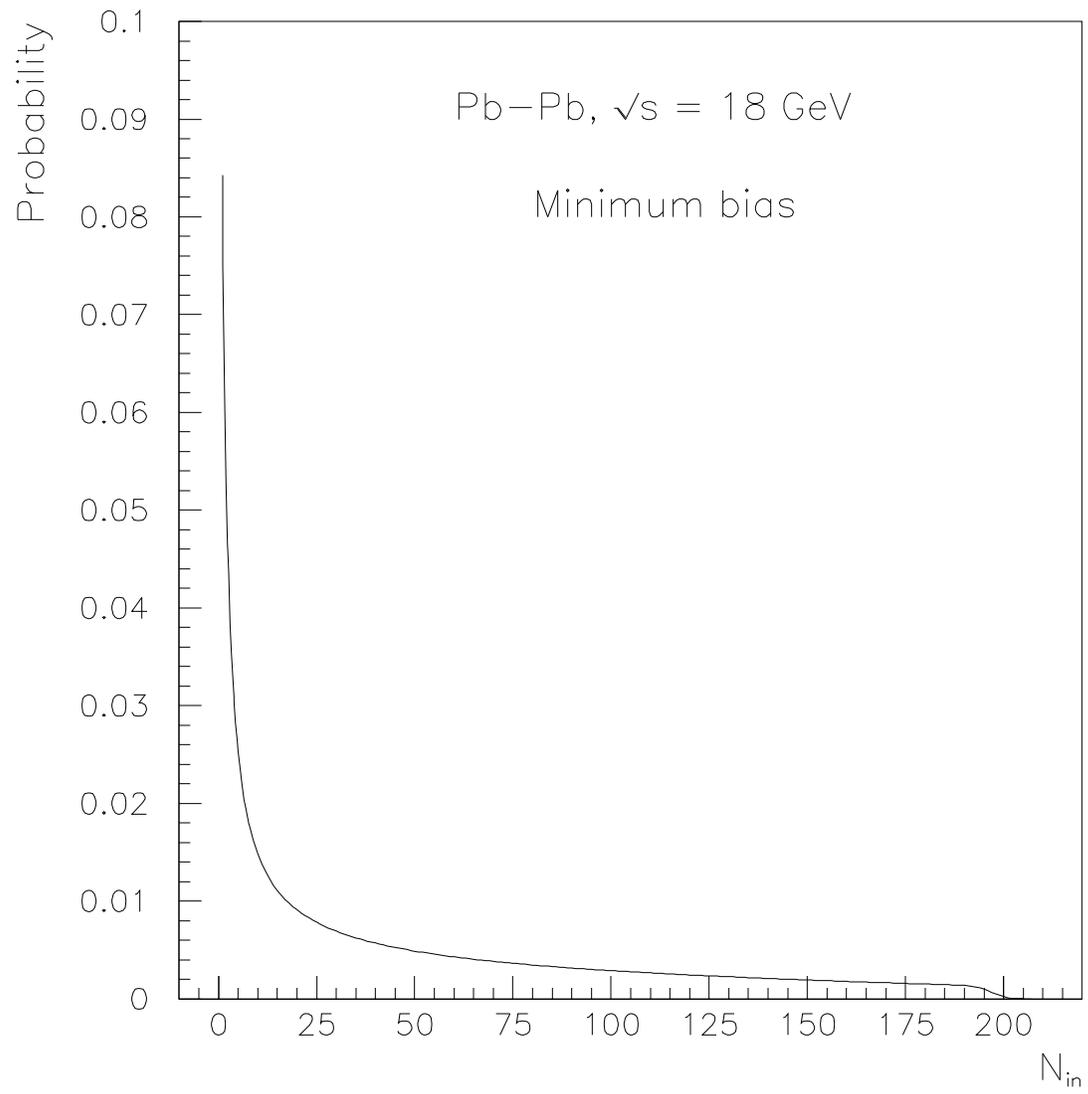
